\documentclass[12pt,preprint]{aastex}

\slugcomment{submitted to the {\it Astrophysical Journal}}
\shorttitle{Population III Enrichment and Preheating of the ICM}
\shortauthors{Loewenstein}

\begin{document}

\title{The Contribution of Population III to the Enrichment and
Preheating of the Intracluster Medium}
\author{Michael Loewenstein\altaffilmark{1}} 
\affil{Laboratory for High Energy Astrophysics, NASA/GSFC, Code 662,
Greenbelt, MD 20771}
\altaffiltext{1}{Also with the University of Maryland Department of
Astronomy} \email{loew@larmes.gsfc.nasa.gov}

\begin{abstract}
Intracluster medium (ICM) abundances are higher than expected assuming
enrichment by supernovae with progenitors belonging to the simple
stellar population (SSP) observed in cluster galaxies, if stars formed
with a standard initial mass function (IMF).  Moreover, new results on
ICM oxygen abundances imply that nucleosynthesis occurred with
nonstandard yields.  The hypothesis that hypernovae (HN) in general,
and HN associated with Population III (Pop III) stars in particular,
may significantly contribute to ICM enrichment is presented and
evaluated. The observed abundance anomalies can be explained by a
hypernovae-producing subpopulation of the SSP, but only if it accounts
for half of all supernova explosions and if Type Ia supernova rates
are very low. Also, the implied energy release may be excessive.
However, an independent Pop III contribution -- in the form of
metal-free, very massive stars that evolve into hypernovae -- can also
account for all the observed abundances, while avoiding these
drawbacks and accommodating a normal IMF in subsequent stellar
generations. The required number of Pop III stars provides sufficient
energy injection (at high redshift) to explain the ICM ``entropy
floor''. Pop III hypernova pre-enrich the intergalactic medium, and
can produce a significant fraction of the metals observed in the
Ly$\alpha$ forest.  Several testable predictions for ICM and IGM
observations are made.
\end{abstract}

\keywords{}

\section{Introduction}

Because rich clusters of galaxies are the largest virialized
structures in the universe, their demographics are useful
discriminants of fundamental cosmological parameters and theories of
large scale structure formation. They also comprise an astrophysical
laboratory for studying physical processes involved in galaxy
formation and evolution: the depths of their dark matter potential
wells imply that -- unlike individual galaxies and perhaps even groups
and poor clusters -- they are good approximations to ``closed boxes'',
and thus ideal sites for investigating the star formation history and
chemical evolution imprinted in the properties of the accumulated
baryonic matter component.  The hot intracluster medium (ICM) is
particularly suitable for such investigations due to the relatively
straightforward measurement of its thermal and chemical properties via
X-ray imaging spectroscopy. This is especially true now, following the
launch of the new generation of X-ray Observatories that includes {\it
Chandra} and {\it XMM-Newton}.

The ICM constitutes a vast reservoir of mass ($\sim 10^{14}$
M$_{\odot}$) and thermal energy ($\sim 3\ 10^{63}$ erg). Yet even
though stars constitute only about a tenth of the baryonic mass,
signatures of the influence of star formation on the ICM are apparent.
Firstly, the ICM is enriched to a significant fraction of the solar
metallicity (e.g., White 2000). The measured amount of metals {\it
cannot} be explained as originating in a simple (coeval, homogeneous)
stellar population (SSP) that also includes the stars observed today
if the stellar initial mass function (IMF) was similar to that in the
solar neighborhood and standard nucleosynthetic yields are assumed
(Loewenstein \& Mushotzky 1996, Section 2).  Secondly, significant
heating of the ICM is evident in departures in the (X-ray)
luminosity-temperature, (total) mass-temperature, and (central ICM)
entropy-temperature relations from predictions of self-similar
(no-heating) scaling (e.g., Loewenstein 2000).  Moreover, by some
accounts, the required heating ($>1$ keV/particle; Wu, Fabian, \&
Nulsen 2000) exceeds even that associated with the number of
supernovae needed to enrich the ICM to its observed metal abundance,
assuming a reasonable energy conversion efficiency. Metallicities and
line widths of Ly$\alpha$ clouds demonstrate that the entire
intergalactic medium (IGM) has been profoundly affected by physical
processes associated with star formation \citep{co,e00,a01,cb}.

Several, seemingly unrelated, recent astrophysical developments may
shed some light on these puzzles, and motivate the present work.  (1)
Spectral analysis of {\it Newton-XMM} data is revealing relative O
abundances well below predictions of standard enrichment theory
\citep{t01, ka01, p01, b01}. (2) Interestingly, calculations of
nucleosynthesis in hypernovae (HN) -- where explosion energies are
10--100 times greater than in standard supernovae -- yield more
extensive oxygen-burning zones and hence depleted O abundances
\citep{n01}.  (3) Furthermore, theoretical arguments now suggest (1)
that the first, metal-free, generation of stars (Population III,
hereafter Pop III) may be predominantly supermassive (Bromm, Coppi, \&
Larson 1999, 2001; Abel, Bryan, \& Norman 2000, Larson 2001) (b) that
such stars may be structurally stable over some mass range
\citep{bhw}, and (c) that these stars may end up exploding as HN with
prodigious production of metals in substantially different relative
proportions than in supernovae with Population I or II progenitors
\citep{h01}. In the following sections, I evaluate the feasibility and
implications of a substantial contribution to the enrichment and
heating of the ICM by HN in general and Pop III HN in particular.  I
adopt the following cosmological parameters: $\Omega_{\rm
matter}=0.3$, $\Lambda=0.7$, and $H_{\rm o}=70$ km s$^{-1}$
Mpc$^{-1}$.

\section{The Enrichment Paradox}

The baryon fraction in a rich cluster of galaxies is $f_{\rm
baryon}=0.155(1+\mu)$, where 0.155 is the measured gas fraction
\citep{l00} and $\mu$ the mass ratio of stars (including remnants of
evolved stars) to gas. Values of $f_{\rm baryon}=0.16-0.20$ for
$\mu\le 0.3$\ can be compared to $0.12-0.15$ obtained using
$\Omega_{baryon}$ inferred from fitting big bang nucleosynthesis (BBN)
models to measured high-redshift Ly-$\alpha$ absorber deuterium
abundances \citep{bnt}.\footnote{The cluster data marginally favors
higher values of $\Omega_{\rm matter}$, or higher $\Omega_{baryon}$ as
inferred from recent cosmic microwave background anisotropy
measurements \citep{hu01}.  Lower values of $\Omega_{baryon}$
consistent with the higher deuterium abundance of one low-redshift
Ly-$\alpha$ absorber and milder lithium depletion requires
$\Omega_{\rm matter}<0.13$ \citep{osw} for consistency with the
cluster data, assuming the latter is ``representative''.}  Consider
chemical enrichment by an SSP, as is appropriate if dominated by
early-type galaxies or their progenitors: most of the stellar mass in
rich clusters resides in elliptical galaxies \citep{a92}.  For the
broken power-law stellar initial mass function that characterizes a
consensus of local IMF (hereafter, LIMF) estimates \citep{kr01}, the
number of Type II supernovae (SNII) explosions per mass of stars
formed is $\approx 0.011$ ${{\rm M}_{\odot}}^{-1}$, assuming all stars
more massive than 8M$_{\odot}$ result in SNII. However for a coeval
stellar population of age comparable to that of the universe, $\approx
40$\% of the original mass will have been shed by stars more massive
than the main sequence turnoff during the course of their evolution
(adopting remnant masses from Ferreras \& Silk 2000). Thus the number
of SNII explosions per solar mass of {\it present-day} main-sequence
stars plus remnants, the specific SNII rate $\eta_{\rm II}\approx
0.019$ ${{\rm M}_{\odot}}^{-1}$ -- a value insensitive to assumptions
about the exact turnoff mass ($0.87{\rm M}_{\odot}$), or upper
($100{\rm M}_{\odot}$) and lower ($0.1{\rm M}_{\odot}$) IMF cutoff
masses.  For IMFs with single slopes of 1.3, 2.0 and, 0.7, $\eta_{\rm
II}=0.013$, $0.0084$, and $0.067$ ${{\rm M}_{\odot}}^{-1}$,
respectively.

To illuminate the enrichment paradox, it is sufficient to consider the
elements O, Si, and Fe. Adopted solar mass fractions, f$_{\odot}$,
\citep{ag} and yields for Type II and Type Ia (SNIa) supernovae
($\langle y_{\rm II}\rangle$ and $y_{\rm Ia}$, respectively; see
Gibson, Loewenstein, \& Mushotzky 1997) are displayed in Table
1.\footnote{SNII yields are averaged over the IMF assuming a slope
$\sim 1.3$ for massive stars; their variation with IMF slope is a
second order effect when compared to differences in $\eta_{\rm II}$.}
I parameterize the contribution of SNIa explosions by the fraction of
cluster baryonic (ICM-plus-stellar) Fe originating in SNIa, $f_{\rm
Ia}{\rm(Fe)}$, that is estimated to be $\sim 0.5$ in the Galaxy
\citep{tww}. The SNIa/SNII ratio, $0.135f_{\rm Ia}{\rm(Fe)}/(1-f_{\rm
Ia}{\rm(Fe)})$, and O/Fe and Si/Fe abundance ratios as functions of
$f_{\rm Ia}{\rm(Fe)}$ are displayed in Table 2 (see, also, Figure 1).
Also shown (highlighted in Table 2) are the values of $\eta_{\rm II}$
needed to reproduce a typical rich cluster ICM Fe abundance of 0.4
solar, the Fe abundance corresponding to the canonical $\eta_{\rm
II}=0.019$ ${{\rm M}_{\odot}}^{-1}$, and the total supernova energy
per unit gas mass, $kT_{\rm SN}$, for both these cases.  Equal
mass-averaged stellar and ICM abundances, $10^{51}$ erg per (Type II
or Ia) supernova, and an ICM-to-stars mass ratio of 10 ($\mu=0.1$) are
assumed.

It is clear that one cannot simultaneously reproduce the observed 0.4
solar Fe abundance and Si/Fe ratio of $\sim 1.7$ \citep{f98} if the
ICM-enriching SNII and current stellar population derive from a single
star formation epoch with the LIMF (see, also, Loewenstein \&
Mushotzky 1996).  Given the observed Fe abundance, the Si abundance
falls short by $\sim 50$\% (see the $f_{\rm Ia}{\rm(Fe)}=0.66$ Table 2
entry and Figure 1) unless the number of SNII is increased by a factor
$\approx 1.8$ ($f_{\rm Ia}{\rm(Fe)}=0.38$ entry) -- which also
increases the supernovae heating by $\sim 60$\%.  An increase in the
average SNII Si yield from 0.14 to 0.24 M$_{\odot}$ is required to
reconcile the observed abundances with the LIMF -- an increase not
supported by any published SNII nucleosynthesis calculations.

Current understanding of star formation is sufficiently incomplete
that the top-heavy or bimodal IMF inferred for the earliest
generations of stars in cluster galaxies (or protogalaxies) by the ICM
observations (see, also, Elbaz, Arnaud, \& Vangioni-Flam 1995;
Matteucci \& Gibson 1995) cannot be ruled out {\it a priori} -- and
may even be supported by other lines of evidence (e.g., Mathews
1989). However, these scenarios are in conflict with the first precise
X-ray measurements of O abundances by the detectors on the {\it
Newton-XMM} Observatory \citep{t01, ka01, p01, b01}. The O/Fe ratio in
the best-fit spectral models for the galaxy clusters Abell 1795, Abell
1835, S\'ersic 159-03, and Virgo ranges from $\sim 0.3-1.0$ solar,
with the lowest values found in the cores of the latter two
systems. If the ICM in these systems -- all of which are cooling flow
clusters -- is intrinsically chemically inhomogeneous, both elemental
abundance and abundance ratio estimates based on relatively simple
models may not accurately reflect the true level and pattern of
enrichment \citep{f01}. However, for the purposes of this paper I
provisionally adopt these values.  Combined with {\it ASCA}
measurements of supersolar Si/Fe ratios \citep{f98}, confirmed with
{\it Newton-XMM} for Virgo \citep{b01}, these indicate a relative
underabundance of O compared to Fe and Si.  The predicted O/Fe ratio
is minimized for high values of $f_{\rm Ia}{\rm(Fe)}$; however, these
produce unacceptably low Si/Fe ratios (Table 2). For example, $f_{\rm
Ia}{\rm(Fe)}=0.85$ implies ${\rm{O/Fe}}\sim 0.5$, but
${\rm{Si/Fe}}\sim 0.7$ (and also $\eta_{\rm II}$ less than half the
LIMF value). Subsolar values of O/Fe imply ${\rm{Si/Fe}}<1.3$ solar.

It is possible that ``standard'' SNII O yields (Table 1) may be
overestimated due to inaccurate reaction rates, an incorrect treatment
of convection, and/or pre-explosion mass loss (Gibson et
al. 1997). The O/Si abundance ratio can be lowered to $\sim 0.5$ by
reducing the assumed IMF-averaged O yield by $\sim 40\%$; although,
the observed Si/Fe ratio still requires an $\sim 80$\% enhancement in
the SNII rate per stellar mass. Here I consider a more radical
alternative -- significant enrichment by hypernovae. My aim is to
determine whether the addition of a HN contribution might account for
these preliminary indications of ${\rm{O/Si}}\sim 0.5$ whilst
maintaining consistency with the observed Si/Fe ratio (for some value
of $f_{\rm Ia}{\rm(Fe)}$); and, if so, to explore the resulting
implications.

\section{Hypernovae, Population III, and the ICM}

\subsection{A Hypernova-Progenitor Subpopulation}

I initially investigate the case of a unimodal IMF where a fraction
$f_{\rm HN}$ of the early-epoch massive stellar population -- perhaps
corresponding to the most metal-poor supernova progenitors -- result
in HN explosions rather than conventional SNII. The closed-box
approximation is used: all nucleosynthetic products from both SNII and
HN are assumed to be retained in the deep cluster potential well.
Nucleosynthetic yields corresponding to the most extreme explosion
kinetic energy studied by \citet{n01}, $10^{53}$ erg, are
considered. Extended burning out to lower density regions results in
relative depletion of O, but enhancement of Si and Fe. For the most
massive He core model they consider, the ratios of yields in the
$10^{53}$ erg model compared to those for $10^{51}$ erg are displayed
in Table 1 ($\epsilon_{\rm HN}$). I adopt these as scaling factors in
estimating the possible contribution of such HN to ICM metal
enrichment.

Table 2 includes two entries for $f_{\rm HN}=0.5$ that is required to
produce ${\rm O/Si}\approx 0.5$. The second, with $f_{\rm
Ia}{\rm(Fe)}=0.23$, is tuned to produce a 0.4 solar Fe abundance
assuming the canonical $\eta_{\rm II}=0.019$ ${{\rm M}_{\odot}}^{-1}$
and yields a Si/Fe ratio about 30\% lower than typically observed in
rich clusters (Figure 1).  The first, with $f_{\rm Ia}{\rm(Fe)}=0$,
simultaneously produces the correct O, Si, and Fe abundances with a
modest increase of $\eta_{\rm II}$ to 0.025 ${{\rm
M}_{\odot}}^{-1}$. The resulting heating is prodigious -- $\sim 25$
keV/particle, equivalent to $\sim 3\ 10^{45}\tau_9^{-1}$ erg s$^{-1}$
per $L_*$ galaxy, compared to $\sim 10^{44}\tau_9^{-1}$ erg s$^{-1}$
per $L_*$ galaxy for $f_{\rm HN}=0$, where the energy is released over
an interval of $10^9\tau_9$ yr. The HN may have more modest energies
$\sim 10^{52}$ erg; but, in that case the yield scaling factors will
be closer to unity and the required value of $f_{\rm HN}$
correspondingly higher.

\subsection{Hypernovae Associated with Population III}

In addition to a parallel subpopulation, I consider a distinct star
formation mode consisting of very massive ($>100$ M$_{\odot}$)
metal-free Pop III stars that ultimately produce ICM-enriching HN.  I
do not address, in detail, the issue of when and how completely the
products of the two star formation modes are ejected from
proto-galaxies; the dispersal of HN products may very well take place
in pre-galactic fragments and the SNII products via subsequent
galactic winds. I also characterize these HN with $f_{\rm HN}$ -- the
fraction of all SNII or HN progenitors that result in hypernovae.  The
absence of metals assures structural stability up until the onset of a
pair-creation instability that proceeds the HN explosion (Baraffe et
al. 2000).  He core masses $M_{\rm core}=100$ and 120 M$_{\odot}$
(from progenitors of mass $\sim 200-250$ M$_{\odot}$), with
corresponding explosion energies 4 and $7\ 10^{52}$ erg, and yield
enhancement factors displayed in Table 1 (as $\epsilon_{\rm HN3}(100)$
and $\epsilon_{\rm HN3}(120)$, respectively) are utilized (adopted
from Heger et al. 2000).  The resulting enrichment and preheating are
shown on the final two lines of Table 2, the former illustrated in
Figure 1.

Intriguingly, unlike all other scenarios described above, the observed
Fe, Si, and O abundances are simultaneously reproduced for a SNII per
stellar mass ratio consistent with the LIMF (Figure 1), if $f_{\rm
HN}=0.005$ and $f_{\rm Ia}{\rm(Fe)}=0.59$ ($M_{\rm core}=100$
M$_{\odot}$), or 0.20 ($M_{\rm core}=120{\rm M}_{\odot}$).  Of the
0.56 (0.58) keV per ICM particle produced by SNII$+$HN for $M_{\rm
core}=100$ (120), 0.08 (0.14) keV per particle originates in the HN
component. This relatively modest energy, however, is released at very
high redshift ($z\gtrsim 10$) where it has maximal effect on the
entropy -- and hence subsequent evolution and final state -- of the
proto-ICM \citep{tn}.

The fraction of the total cluster baryonic mass originating as Pop III
HN progenitors is
\begin{equation}
{{M_{\rm HNP}}\over {M_{\rm ICM}+M_{\rm stars}}}=0.019 \left({{f_{\rm
HN}}\over {0.005}}\right) \left({{\eta_{\rm II}}\over {0.019{{\rm
M}_{\odot}}^{-1}}}\right) \left({{M_{\rm prog}}\over {200{\rm
M}_{\odot}}}\right) {{\mu}\over {1+\mu}},
\end{equation}
where $M_{\rm prog}$ is the mean HN progenitor mass, and $\mu$ is the
cluster ratio of stars-to-gas as previously defined.  That is, the
mass density of Pop III stars in units of the critical density,
\begin{equation}
\Omega_{\rm III}=7.1\ 10^{-5}
\left({{\Omega_{\rm baryon}}\over {0.041}}\right)c^{-1}{b_1}^{-1}{b_2}^{-1},
\end{equation}
for $f_{\rm HN}=0.005$, $\eta_{\rm II}=0.019$ ${{\rm
M}_{\odot}}^{-1}$, $M_{\rm prog}=200{\rm M}_{\odot}$, and $\mu=0.1$.
$\Omega_{\rm baryon}$ is normalized to the BBN value, $c$ $(<1)$ is
the fraction of Pop III stars that produce HN and $b_1$ and $b_2$ are
cluster ``bias'' factors encompassing any over- or under-concentration
of baryons and any relative Pop III stellar formation probability
enhancement in clusters relative to the universal average,
respectively.  \citet{og} estimated $\Omega_{\rm III}$ from
self-consistent, semi-analytic modeling of the evolution of the Jeans
mass through the epoch of IGM reheating. Their results imply
$b_2\approx 17$ (for $c\sim 1$, $b_1\sim 1$). A value $b_2>>1$ is not
unexpected: Pop III stars are unlikely to have formed as readily (or,
perhaps, as early or with the same IMF) outside of these regions of
highest primordial overdensity. A low value of $\Omega_{\rm baryon}$
(as, e.g., implied by a high primoridal deuterium abundance) would
indicate $b_1>1$ and a correspondingly lower value of $b_2$.

\section{Discussion and Predictions}

\subsection{Summary of Hypernovae ICM Enrichment}

If the stars responsible for enriching the ICM and the stars observed
today in cluster early-type galaxies originate in the same early star
formation epoch, then the amount of Si observed in clusters cannot be
explained with standard SNII yields and the solar neighborhood IMF:
the number of SNII per stellar mass must be increased by nearly a
factor of two through invocation of a flat or bimodal IMF. However,
recent observations of the ICM ratio of O/Si that are well below solar
cast doubt on this explanation.  For reasonable SNIa/SNII ratios, Si
and O are predominantly synthesized in SNII that produce a roughly
solar O/Si ratio.

Although HN nucleosynthesis calculations are at an early stage,
depleted O abundances caused by more extensive O-burning are likely to
persist. Thus, it is worth studying their possible role in ICM
enrichment -- particularly in light of their potential to preheat the
ICM and their possible connection to the elusive Population III stars.
I investigate two classes of HN contributions. Firstly, I consider the
case where some fraction (perhaps the most metal-poor) of SNII
progenitors from a unimodal IMF give rise to hypernovae with increased
explosion energy and nonstandard yields, as in the calculations of
\citet{n01}.  Their contribution can indeed lower the O/Si ratio to
the observed level, but only if there are approximately equal numbers
of HN and conventional SNII. Moreover, this implies an energy
production of $>20$ keV per ICM particle -- sufficient to unbind the
ICM of even the most massive clusters unless most of the $\sim
10^{64}$ erg of energy is radiated -- as well as negligible enrichment
by SNIa.
 
ICM abundances are more naturally explained with a substantial
contribution from Population III hypernovae originating in an earlier,
independent star formation mode. An absence of metals reduces cooling
and leads to a very large Jeans mass: Pop III stars are likely to form
with an extremely top-heavy IMF \citep{l01}.  (A hybrid scenario is
also possible if the Pop III IMF is itself bimodal; Nakamura \&
Umemura 2001.)  Very massive, metal-free stars naturally give rise to
hypernovae with enhanced yields and skewed abundance ratios compared
to ordinary SNII. Utilizing a pair of representative cases from
\citet{h01}, I find that if one such HN contributes to ICM enrichment
for every 200 SNII, the observed {\it proportions} of Fe, Si, and O
are simultaneously explained. Also, the additional contribution from
ordinary SNII required to explain the {\it amount} of these elements
is consistent with the LIMF, in keeping with evidence for a universal
IMF \citep{w97}; this is not true of an alternative scenario for
simultaneously explaining the relative amounts of O, Si, and Fe by
reducing SNII O yields.  Finally, the associated preheating of $\sim
0.1$ keV per particle is sufficient to account for the observed
cluster ``entropy floor'' (e.g., Lloyd-Davies, Ponman, \& Cannon
2000), since it is deposited at $z\gtrsim 10$ when the more diffuse
ICM was especially susceptible to a shift to a high adiabat
\citep{tn}.

Under this scenario, cluster O primarily (90\%) originates from SNII,
while comparable contributions from SNII and HN account for Si.  About
one-third of the ICM Fe originates from SNII, with HN contributing
some amount $\le$ one-half (this is uncertain due to the steep
dependence of HN Fe yields on progenitor core mass) and SNIa the rest.
The partial decoupling of the origins of these elements has
implications for expectations of future accurate X-ray measurements of
ICM abundances, in addition to the prediction that subsolar O/Si
ratios will be confirmed and found to be common.  Correlations of the
mass in each of these metals versus optical light may have offset
zero-points relative to the case where all metals share a monolithic
origin.  Cluster-to-cluster variations in metal mass-to-optical-light
ratio may display an elemental dependence with, e.g., ${\rm M_{\rm
Si}}/{\rm L}$ displaying a greater scatter than ${\rm M_{\rm O}}/{\rm
L}$. Si abundances are predicted to be substantial even at very high
redshift, and evolve more slowly than O abundances.
 
\subsection{Implications for the Enrichment of the IGM}

In current models for the chemical evolution of the IGM, the
enrichment is dominated by supernovae associated with early-epoch,
Population II star formation. Yields of $\sim 2-5\times$ the solar
mass fraction of metals for each solar mass of star formation are
typically required to match the C abundances measured in Ly$\alpha$
forest clouds \citep{g98,co,a01,cb}. Population III stars have been
proposed as providing the radiation that reionizes the universe
\citep{og,ts}, while simultaneously pre-enriching the IGM to modest
levels $\le 10^{-3}$ solar \citep{og,wq}. However, if hypernova are
produced by Population III as suggested here in order to explain ICM
abundances, the implied pre-enrichment may be significantly
enhanced. For the value of $\Omega_{\rm III}$ calculated by \citet{og}
(corresponding to a Pop III stellar mass per baryon ratio 17 times
less than proposed here for the ICM) the \citet{h01} HN yields imply
IGM pre-enrichment to as much as $\sim 2.5\ 10^{-3}$ solar metallicity
with very non-solar abundance ratios.  While O (and elements of lesser
atomic weight) are synthesized in amounts small compared to those
observed in Ly$\alpha$ forest clouds, Si attains $>10^{-2}$ solar
abundance at $z\sim 10$. Since subsequent Pop II enrichment must then
be invoked to produce $\sim 10^{-2}$ solar C abundances, a Si/C ratio
about twice solar is expected. This is consistent with measured UV
line ratios \citep{gs,s98} in the clouds. Thus a contribution to IGM
enrichment from Pop III hypernovae can alleviate the requirement of
excessive Population II yields, while explaining a likely observed
overabundance of Si. Similar overabundances of, e.g., S and Ca -- but
not N or O -- relative to C are predicted -- the former being produced
in roughly equal amounts by Populations II and III, the latter
primarily by Population II.

\section{Concluding Remarks}

Signatures of hypernova explosions of very massive Population III
stars are found in the numbers and abundance pattern of
low-metallicity Milky Way stars \citep{fh,un} and in the population of
$>100$ M$_{\odot}$ black holes \citep{mr}.  I have demonstrated that
these extreme primordial events also leave a thermal and chemical
imprint on the IGM and ICM.  Direct observation of hypernova
explosions and/or their progenitors with the {\it Next Generation
Space Telescope}, in concert with advances in calculations of HN
yields, could provide direct confirmation of the important role of Pop
III in enriching both the IGM and ICM.  Population III hypernovae may
constitute an important feedback mechanism during the earliest galaxy
formation era and ought to be be considered in semi-analytic galaxy
formation calculations.

\acknowledgments

I am grateful for feedback from the referee and from Richard
Mushotzky.

\clearpage


\clearpage

\begin{deluxetable}{cccccccc}
\tablecaption{SN Yields and HN Enhancement Factors}
\tablewidth{0pt}
\tablehead{
\colhead{} & \colhead{f$_{\odot}$} & \colhead{$y_{\rm Ia}$} & 
\colhead{$\langle y_{\rm II}\rangle$} & \colhead{$\epsilon_{\rm HN}$} & 
\colhead{$\epsilon_{\rm HN3}(100)$} & \colhead{$\epsilon_{\rm HN3}(120)$}   
}
\startdata
0 & $9.6\ 10^{-3}$ & 0.15 & 1.7 & 0.68 & 25 & 20 \\
Si & $7.1\ 10^{-4}$ & 0.16 & 0.14 & 1.9 & 160 & 180 \\
Fe & $1.3\ 10^{-3}$ & 0.74 & 0.10 & 3.5 & 50 & 270 \\
\enddata

\tablecomments{Yields are in M$_{\odot}$.}

\end{deluxetable}

\clearpage

\begin{deluxetable}{cccccccccc}
\tablecaption{Results for Various Supernova Combinations}
\tablewidth{0pt}
\tablehead{
\colhead{$f_{\rm Ia}{\rm(Fe)}$} & \colhead{f$_{\rm HN}$} & 
\colhead{Ia/II} & \colhead{O/Fe} & \colhead{Si/Fe} & 
\colhead{$\eta_{\rm II}$\tablenotemark{a}} & 
\colhead{$kT_{\rm SN}$\tablenotemark{a}} & \colhead{Fe\tablenotemark{b}} & 
\colhead{$kT_{\rm SN}$\tablenotemark{b}}
}
\startdata
0 & 0 & 0 & 2.2 & 2.5 & 0.056 & 1.2 & 0.14 & 0.40 \\
0.25 & 0 & 0.045 & 1.7 & 2.0 & 0.042 & 0.93 & 0.18 & 0.42 \\ 
{\bf 0.38} & {\bf 0} & {\bf 0.083} & {\bf 1.4} & {\bf 1.7}
           & {\bf 0.035} & {\bf 0.80} & 0.22 & 0.44 \\ 
0.5 & 0 & 0.135 & 1.1 & 1.4 & 0.028 & 0.67 & 0.27 & 0.46 \\
{\bf 0.66} & {\bf 0} & {\bf 0.26} & {\bf 0.78} & {\bf 1.1}  
           & {\bf 0.019} & {\bf 0.51} & {\bf 0.4} & {\bf 0.51} \\
0.75 & 0 & 0.405 & 0.58 & 0.91 & 0.014 & 0.42 & 0.54 & 0.57 \\
{\bf 0} & {\bf 0.5\tablenotemark{c}} & {\bf 0} & {\bf 0.83} & {\bf 1.6}
        & {\bf 0.025} & {\bf 26} & 0.31 & 20 \\
{\bf 0.23} & {\bf 0.5\tablenotemark{c}} & {\bf 0.092} & {\bf 0.65} & {\bf 1.3}
           & {\bf 0.019} & {\bf 20} & {\bf 0.4} & {\bf 20} \\
{\bf 0.59} & {\bf 0.005\tablenotemark{d}} & {\bf 0.24} & {\bf 0.84} & 
             {\bf 1.7} & {\bf 0.018} & {\bf 0.56} & {\bf 0.41} & {\bf 0.58} \\
{\bf 0.20} & {\bf 0.005\tablenotemark{e}} & {\bf 0.079} & {\bf 0.84} & 
             {\bf 1.7} & {\bf 0.019} & {\bf 0.58} & {\bf 0.4} & {\bf 0.57} \\
\enddata

\tablecomments{Fe abundances and abundance ratios are relative to
solar, $kT_{\rm SN}$ is in keV, $\eta_{\rm II}$ in ${{\rm
M}_{\odot}}^{-1}$.}  \tablenotetext{a}{values corresponding to solar
Fe abundance}
\tablenotetext{b}{values corresponding to Local IMF value of
$\eta_{\rm II}$}
\tablenotetext{c}{hypernovae as in \citet{n01}}
\tablenotetext{d}{hypernovae as in \citet{h01}, $M_{\rm core}=100$}
\tablenotetext{e}{hypernovae as in \citet{h01}, $M_{\rm core}=120$}

\end{deluxetable}

\clearpage


\begin{figure}
\plotone{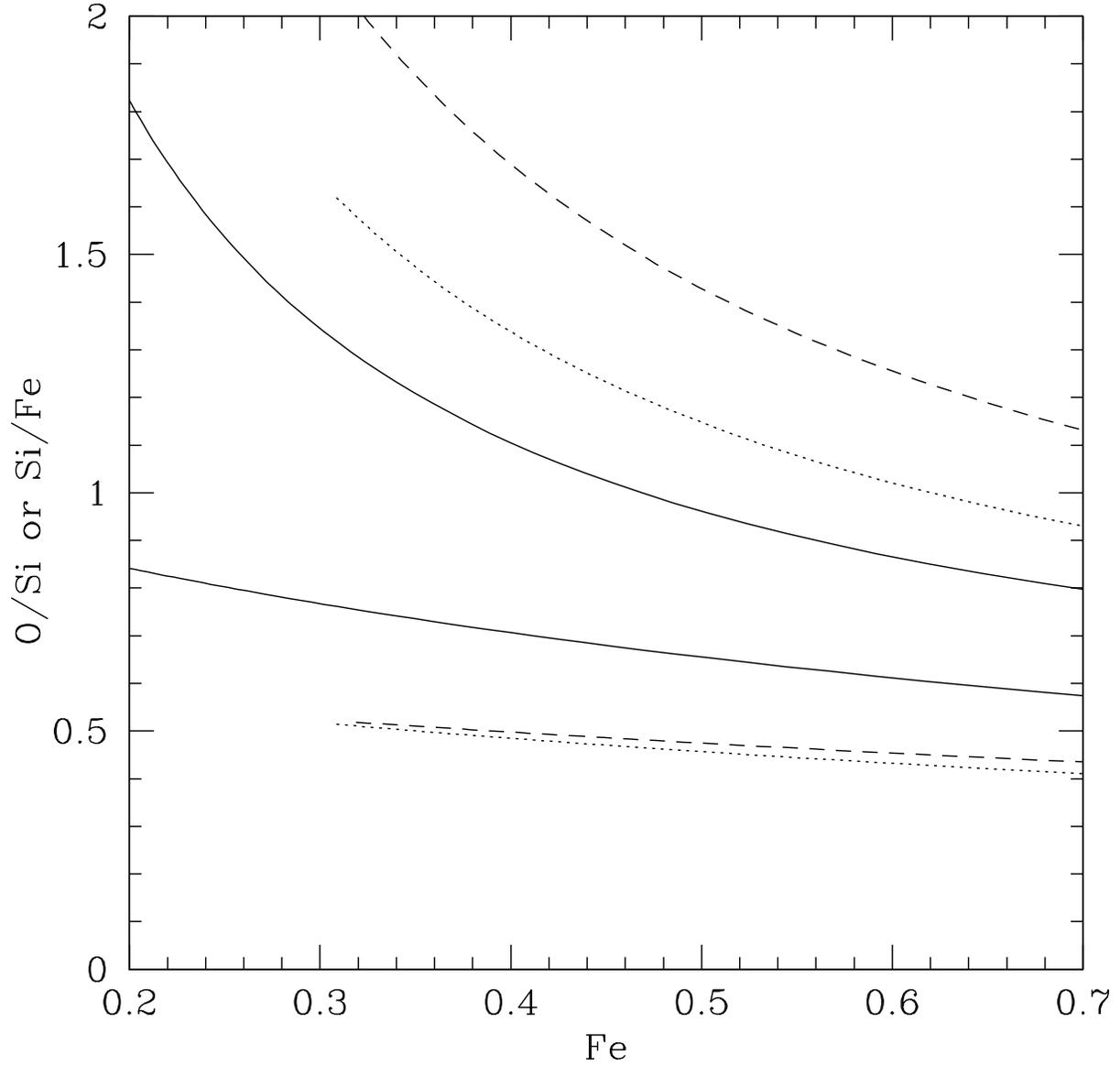}
\caption{O/Si (lower curves) and Si/Fe (upper curves) abundance
ratios as a function of Fe abundance for the LIMF specific SNII
rate. Solid curves illustrate the case of no hypernovae, dotted curves
hypernovae as in \citet{n01}, dashed curves hypernovae as in
\citet{h01}. \label{fig1}}
\end{figure}

\clearpage 

\end{document}